\documentclass[10pt, conference]{IEEEtran}
\usepackage{amsfonts}
\usepackage{amssymb}
\usepackage{eurosym}
\usepackage{cite}
\usepackage{graphicx}
\usepackage{epstopdf}
\usepackage{amsmath}
\usepackage{tikz,lipsum}
\usepackage{caption}
\usepackage[T1]{fontenc}
\usepackage{amsthm}
\usepackage{mathrsfs}
\usepackage{color}
\usepackage{stfloats}
\usepackage{xcolor, etoolbox}
\usepackage{subcaption}
\usepackage{float}
\usepackage{authblk}
\usepackage{placeins}

\IEEEoverridecommandlockouts
\makeatletter
\def\ps@IEEEtitlepagestyle{%
    \def\@oddhead{\parbox{\textwidth}{\centering\footnotesize This is the authors' original preprint version of a paper accepted for publication at the 8th International Conference on Advanced Communication Technologies and Networking (CommNet 2025), Rabat, Morocco, December 3--5, 2025. The final version will be published and made available via IEEE Xplore.}\hfill}%
    \def\@evenhead{}%
    \def\@oddfoot{\parbox{\textwidth}{\centering\footnotesize This is the preprint version of a paper accepted for publication at CommNet 2025. The final version will appear in the official IEEE proceedings}\hfill}%
    \def\@evenfoot{}%
}
\makeatother

\def\BibTeX{{\rm B\kern-.05em{\sc i\kern-.025em b}\kern-.08em
    T\kern-.1667em\lower.7ex\hbox{E}\kern-.125emX}}
\begin{document}

\title{Bin2Vec: Interpretable and Auditable Multi-View Binary Analysis for Code Plagiarism Detection}
\author[1]{Moussa MOUSSAOUI}
\author[2]{Tarik HOUICHIME}
\author[3]{Abdelalim SADIQ}
\affil[1]{\textsuperscript{,3}Computer Science Research Laboratory(LaRI), Faculty of Science, Ibn Tofail University,\protect\\ Kenitra, 14000, Morocco}
\affil[2]{LRIT, Faculty of Science, Mohammed V University In Rabat, Rabat, 10112, Morocco}

\affil[ ]{Emails: $^{1}$\,moussa.moussaoui@uit.ac.ma, $^{2}$\,tarik\_houichime@um5.ac.ma, $^{3}$\,a.sadiq@uit.ac.ma}
\maketitle

% tighten vertical space around floats to keep sequential figures compact
\setlength{\textfloatsep}{8pt plus 1pt minus 2pt}
\setlength{\floatsep}{8pt plus 1pt minus 2pt}
\setlength{\intextsep}{8pt plus 1pt minus 2pt}
\setlength{\abovecaptionskip}{6pt plus 1pt minus 2pt}
\setlength{\belowcaptionskip}{4pt plus 1pt minus 2pt}

% Abstract (wrapped per template structure)
\begin{abstract}
We introduce Bin2Vec, a new framework that helps compare software programs in a clear and explainable way. Instead of focusing only on one type of information, Bin2Vec combines what a program looks like (its built-in functions, imports, and exports) with how it behaves when it runs (its instructions and memory usage). This gives a more complete picture when deciding whether two programs are similar or not. Bin2Vec represents these different types of information as views that can be inspected separately using easy-to-read charts, and then brings them together into an overall similarity score. Bin2Vec acts as a bridge between binary representations and machine learning techniques by generating feature representations that can be efficiently processed by machine-learning models. We tested Bin2Vec on multiple versions of two well-known Windows programs, PuTTY and 7-Zip. The primary results strongly confirmed that our method compute an optimal and visualization-friendly representation of the analyzed software. For example, PuTTY versions showed more complex behavior and memory activity, while 7-Zip versions focused more on performance-related patterns. Overall, Bin2Vec provides decisions that are both reliable and explainable to humans. Because it is modular and easy to extend, it can be applied to tasks like auditing, verifying software origins, or quickly screening large numbers of programs in cybersecurity and reverse-engineering work.
\end{abstract}

% Keywords (template position after abstract)

% Sections
\section{Introduction}
Binary code similarity (BCS) forms the base for various critical tasks,  such as hunting security flaws and examining software patches, tracing  malware origins and grouping them, detecting code sources or plagiarism  in multiple releases, and scanning vast code repositories. Though highly useful, BCS remains inherently challenging: compilers and linkers  modify instruction and data flows; removing identifiers eliminates  meaningful indicators of code intent; deliberate obfuscation distorts  the overall structure; and comparisons across diverse architectures or  optimization variants produce significant structural variations. In  practical settings, these approaches must expand from identifying  differences in small code segments to retrieving matches over entire  programs within hundreds of thousands of artifacts.

Graph-based methodologies translate code functions into semantic vector encodings, enabling robust similarity detection despite compiler-induced variations.
While this structural-semantic approach captures deep programmatic logic, it is computationally intensive and struggles with significant code obfuscation or cross-architectural comparisons. An alternative, high-performance approach scales function-level vector encodings for rapid, real-time matching, trading explanatory power for build resilience and speed.
Massive-scale comparisons are achieved by further compressing entire programs into compact, fixed-size representations, which introduces a trade-off between search velocity and the potential for detail loss.

These static alignment techniques optimize for cross-variant matching speed and quality, but they remain vulnerable to structural obfuscation and are fundamentally incapable of capturing dynamic runtime behavior.

To address this gap, we introduced Bin2Vec (Binary to Vectors), a modular trace-centric methodology. Which constructs compact, comparable embeddings from multiple complementary signals that are readily available or derivable in practical reverse engineering work-flows. We capture the dynamic behaviour using x64dbg alongside static informations from Ghidra \cite{b74, b73}. This multi-view design aims to be robust to single-signal brittleness while maintaining interpretability. We also provide a reproducible end-to-end pipeline that curates and merges traces, derives per-signal embeddings with deterministic fallback, computes per-feature and global similarities.  By explicitly incorporating behavioral analysis via execution traces and register usage, Bin2Vec complements graph and hashing-based approaches \cite{b10, b15, b38, b62, b71}. Its practical advantages are transparency, resilience to any single broken signal.

\section{Related Work}
Binary comparison has progressed from older methods that mainly looked at program structure to newer approaches that use machine learning. Early tools, such as BinDiff \cite{b66}, compared programs by checking how their parts were connected and labeled, which made it useful for spotting changes between software versions. Later, tools such as BinHunt went further by analyzing how information flows inside the program, helping to detect deeper differences \cite{b43}. While these tools worked well for tracking versions and helping experts analyze software, they mostly compared code piece by piece and could struggle when the software was heavily modified, disguised, or reorganized.

Machine learning has expanded how programs can be compared by turning pieces of code into numerical representations and then using them to make broader program-level decisions. Some approaches focus on the structure and meaning of code, using graphs that show how different parts depend on each other or how the program flows \cite{b59, b10}. Others train models to recognize differences and similarities between functions, or combine multiple views of a program to make the comparison more reliable across different versions \cite{b31}, \cite{b34}, \cite{b35}, \cite{b36}. Researchers have also experimented with techniques inspired by language translation, treating code like a language to be modeled \cite{b31}. Overall, machine learning has improved the accuracy of binary comparison, but challenges remain, especially when it comes to handling very large programs, explaining results clearly, and making sure methods work reliably across many kinds of software variations \cite{b43, b39}.

The State-of-the-art of BCS is advancing on two complementary fronts.  First, methods based on graphs represent low-level code functions as visual  diagrams and create vector  encodings that capture the code's semantics to match similar functions  \cite{b59}. For example, CSGraph2Vec generates distributed,  graph-inspired representations of these code functions that detect  structural patterns and contextual details, supporting effective similarity searches or classifications, even across code variations from different compilers or optimization settings  \cite{b10}.

ReGraph accelerates cross-platform function alignment \cite{b62} and excels in speed and cross-variant robustness. It relies on clean  graphs, so errors in disassembly or function boundary detection can  cascade, and it’s less resilient to heavy obfuscation that scrambles  graph structure. Concurrently, BinEGA \cite{b71} presents an efficient graph alignment method that  enhances DNN-based BCS pipelines, achieving better match quality at  lower runtime vital for large-scale deployments. It delivers better matches under common compiler tweaks, but heavy code  reshuffling can throw it off, and memory use and run time can still grow as artifacs get larger. These systems mostly care about shape, not behavior: they compare code  without watching it run, leaving space for runtime‑driven matching. A major advantage is its capacity to capture detailed structural  meanings that go beyond simple sequences of code elements. However, a  common drawback is the reliance on precisely reconstructing these  diagrams (such as function outlines and flowcharts for control or data  paths), along with the high computing demands of extracting and  comparing them, which can falter when the code is heavily disguised or  when adapting it across different hardware architectures. Second,  researchers have investigated expanding these learned representations to search at the level of entire programs.

BinSAFE examines methods for scaling up intelligent vector encodings of code  functions to spot similarities in compiled code, prioritizing quick  performance and reliability no matter how the code was built \cite{b15}.  Its key benefits are efficient real-time handling and stronger  resilience to changes in builds, though the streamlined approach can  lower how easy it is to interpret outcomes and could lead to problems if the training environment shifts or if compilers act in unexpected ways.

KEENHash takes this even further by converting entire programs into  compact, fixed-size representations that are sensitive to individual  functions, using vectors for those functions, which are then grouped and compressed through  clustering and hashing techniques. This allows for comparisons on a  massive scale up to billions while still delivering strong results  \cite{b38, b72}. The biggest upside is its incredible  scalability, with speedy organization and lookup; the downside is the  risk of losing some details through hashing and relying heavily on the quality  of the initial generated features.

Surveys of the field point out common challenges, such as the heavy reliance on precise program analysis tools, the difficulty of dealing with code that has been deliberately disguised or heavily optimized by compilers, and the limited transparency of machine learning models \cite{b39, b58}. Researchers have also shown that deep learning–based systems can sometimes be tricked by specially crafted code, highlighting the need for more robust methods that combine multiple kinds of evidence \cite{b57}. In contrast to methods that only focus on program structure or reduce software to a single, opaque fingerprint, our approach compares programs at scale while keeping the process transparent and interpretable. We combine multiple types of evidence such as functions, imports and exports, execution traces, and register usage collected with widely available tools like Ghidra and x64dbg \cite{b73, b74}. This design allows us to maintain efficiency while also providing clearer insights.
\section{Methodology}
\begin{figure*}[t]
  \centering
  \includegraphics[width=0.95\textwidth]{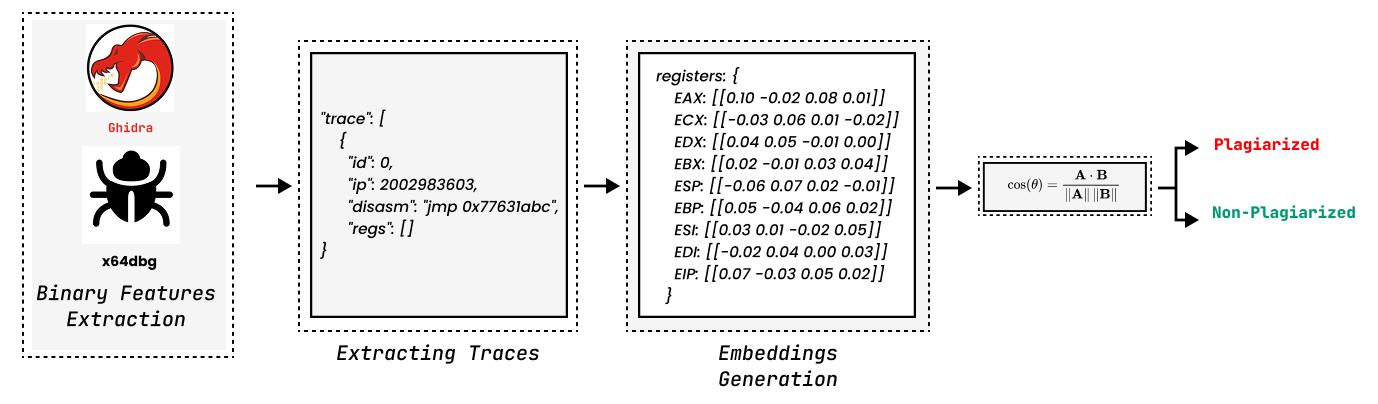}
  \caption{Overview of the Bin2Vec pipeline: inputs (static and dynamic), per-view embedding construction, normalization and aggregation, and global cosine similarity.}
  \label{fig:methodology}
\end{figure*}

Our method Fig.~\ref{fig:methodology} seeks to optimize the representation of software binaries to enable machine learning to easily detect similarities. To show how this works, we compute the representation for two PE artifacts. The process begins with collecting both static and dynamic artifacts. Static information, such as functions, imports, and exports, is extracted using Ghidra, while dynamic information, including execution traces and register usage, is collected with x64dbg \cite{b73, b74}. Together, these complementary inputs form the basis for multiple evidence “views” of a binary representation. Cosine similarity is adopted due established rule for measuring semantic or structural similarity. Since artifacts may target different architectures, we first inspect the PE header to determine whether a sample is 32-bit (PE32) or 64-bit (PE32+). This ensures that subsequent analysis remains architecture-consistent. For dynamic analysis, we run x64dbg in the appropriate mode, producing \texttt{.trace32} or \texttt{.trace64} files accordingly. To make register usage comparable across instruction sets, we normalize register families so that, for example, \{EAX, AX, AL\} in 32-bit and \{RAX, EAX, AX, AL\} in 64-bit both map to a common accumulator dimension. This alignment allows the comparison of register-based features across different builds without losing fidelity.

Once architecture alignment is established, we acquire and merge the static and dynamic information into a unified representation per artifact. Provenance tags are retained so that each feature can be traced back to its original source during later inspection. This combined dataset serves as the input to the embedding construction stage.

The embedding design covers five distinct views, each intended to capture complementary signals. All tokenizable fields are embedded using the \texttt{all-MiniLM-L6-v2} encoder \cite{b18}, with a seeded hashing encoder as fallback to preserve determinism. Function metadata from Ghidra is summarized into descriptors such as parameter count, calling conventions, call connectivity, and size, which are aggregated into a per-artifact vector. Imports and exports are described through their names, libraries, addresses, and categorical flags, and pooled into vectors representing external program interfaces. Dynamic traces from x64dbg are transformed into opcode and instruction token streams, supplemented by activity summaries (e.g., frequency of memory or control-flow operations); these are encoded through bag-of-ops or token n-grams, optionally reduced via PCA for stability. Finally, register usage is summarized by update frequency, read/write ratios, and execution context, with architecture-specific registers mapped to canonical families to allow 32/64-bit comparability.

To ensure consistency across different views, all feature vectors are standardized and normalized. For sets of elements such as multiple functions or imports, we aggregate by mean pooling, yielding one vector per view and per artifact. Feature-wise cosine similarity is then computed between two artifacts for each view, producing a set of similarity scores that reflect agreement across different characteristics. For the overall program-level decision, these per-view vectors are concatenated in a fixed order (functions, imports, exports, traces, registers), and cosine similarity is computed over the concatenated representation. This follows scalable whole-binary retrieval strategies that aggregate function- or feature-level signals \cite{b15, b38}. In addition to the global similarity score, we also report the weighted mean of per-view cosine scores, which provides insight into agreement or disagreement across modalities. Both raw and PCA-reduced variants are maintained to improve stability in the presence of small variations in execution traces. Finally, we highlight several quality and scalability considerations. The pipeline is explicitly architecture-aware, ensuring 32/64-bit comparability; deterministic encoding is guaranteed by the seeded fallback encoder; and cosine similarity provides a simple, transparent scoring rule. Scalability is achieved through fixed-length per-view vectors that compose linearly, allowing whole-artifact comparison in line with recent program-level systems. Optional dimensionality reduction further stabilizes comparisons while keeping the approach lightweight.

\section{Experimental Results and Analysis}

\begin{figure}[!ht]
  \centering
  \includegraphics[width=0.9\linewidth]{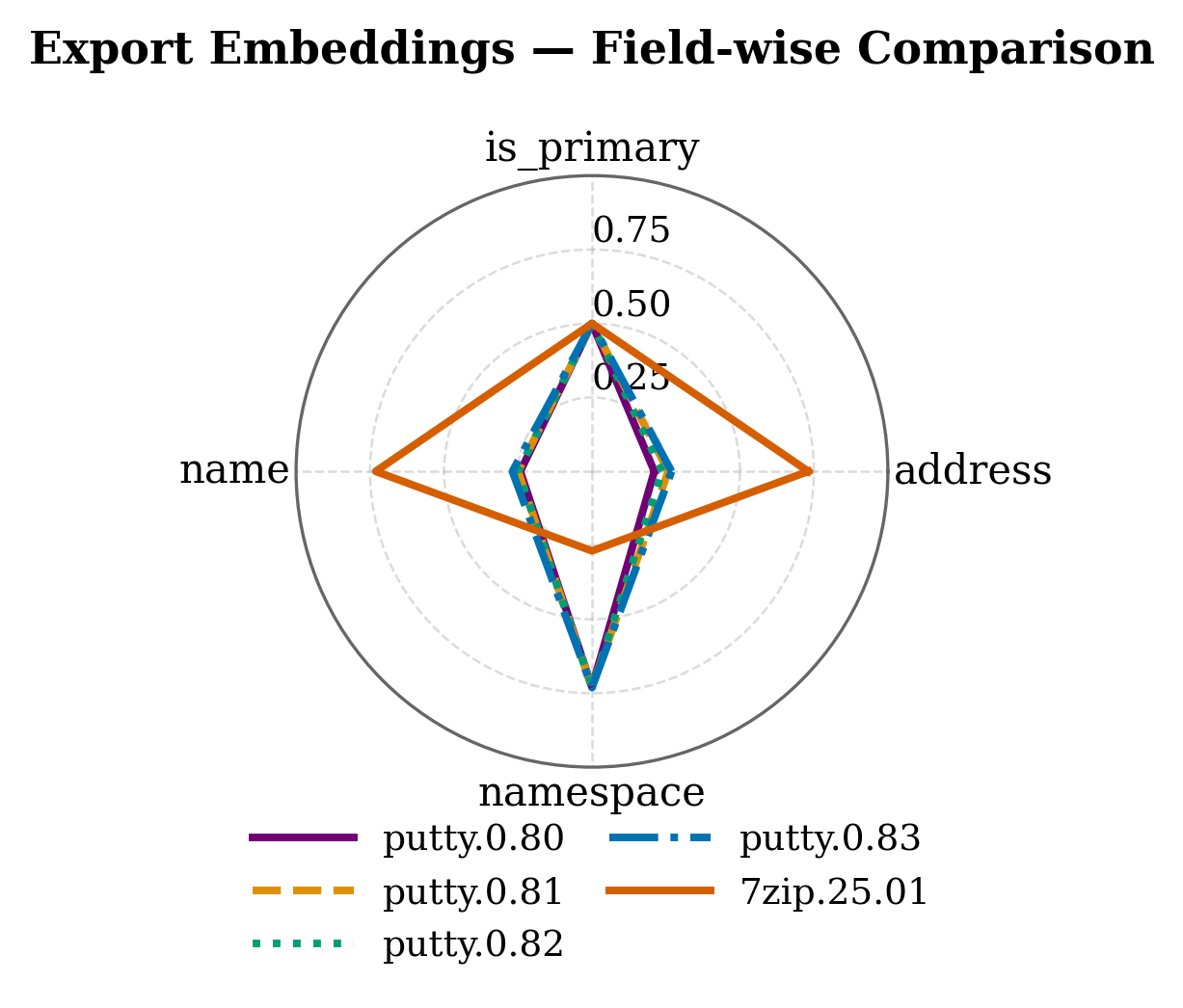}
  \caption{Field-wise export embeddings comparison (cosine similarity): PuTTY (4 versions) vs. 7-Zip.}
  \label{fig:radar-exports}
\end{figure}

We evaluate our approach on multiple Windows programs compiled for PE32: diffrent releases of PuTTY and 7-Zip 25.01, all 32-bit. We extract static views (functions, imports, exports) using Ghidra, while we collect dynamic views (traces and register summaries) with x64dbg. We then merge both static and dynamic information into a unified representation. We produce embeddings deterministically, and when model artifacts are unavailable, we fall back to a seeded hashing encoder to ensure reproducibility. Turning to the comparative analysis, we first examine field-wise results as the initial layer of evidence. Starting with exports, we observe that 7-Zip leans heavily toward address and plain symbol names, whereas PuTTY versions emphasize namespace, suggesting richer decoration or analyzer-inferred grouping Fig.~\ref{fig:radar-exports}; the is\_primary signal remains balanced across all binaries.

\begin{figure}[!ht]
  \centering
  \includegraphics[width=0.9\linewidth]{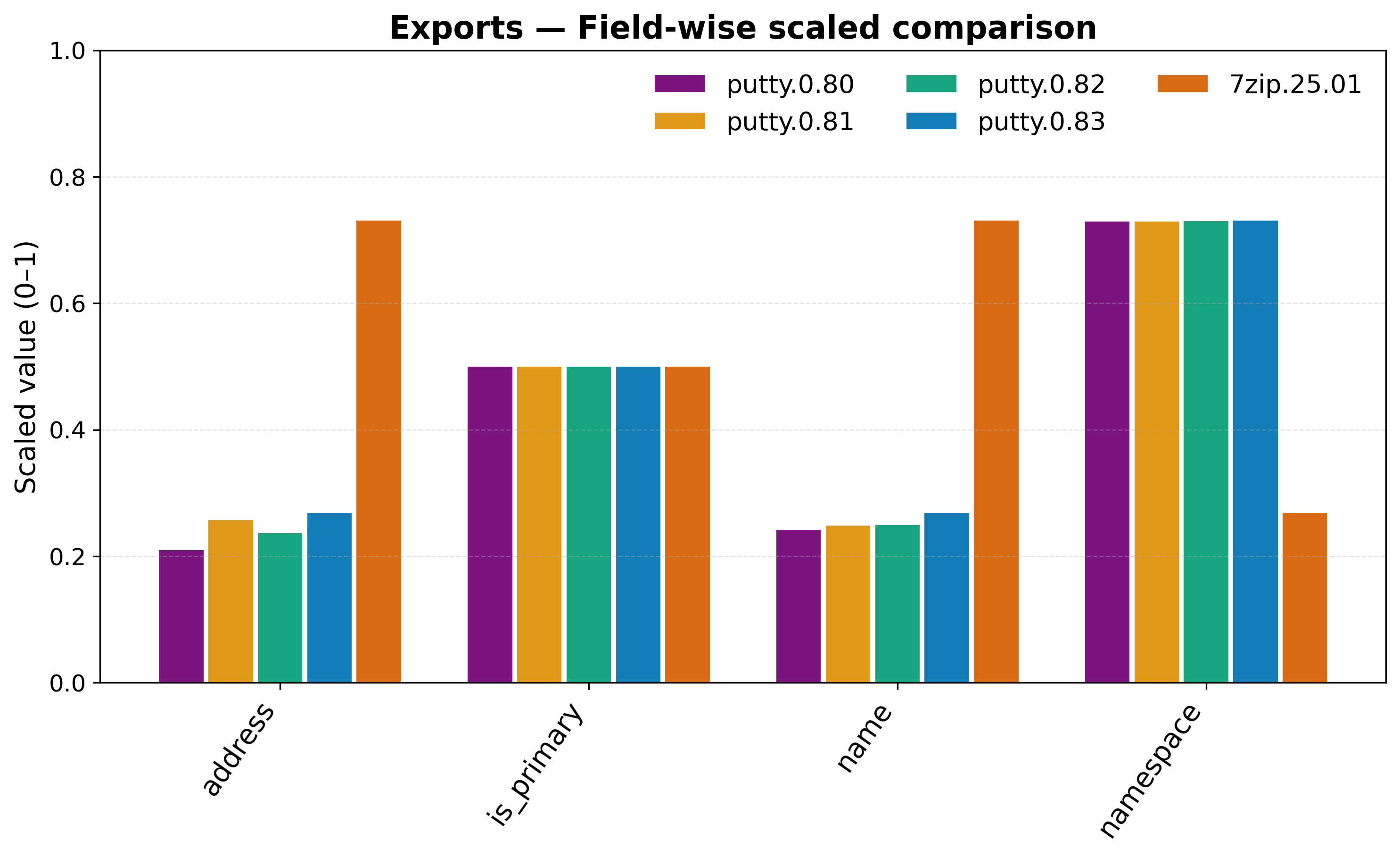}
  \caption{Field-wise export comparison: PuTTY (4 versions) vs. 7-Zip.}
  \label{fig:bars-exports}
\end{figure}

\begin{figure}[!ht]
	\centering
	\includegraphics[width=0.9\linewidth]{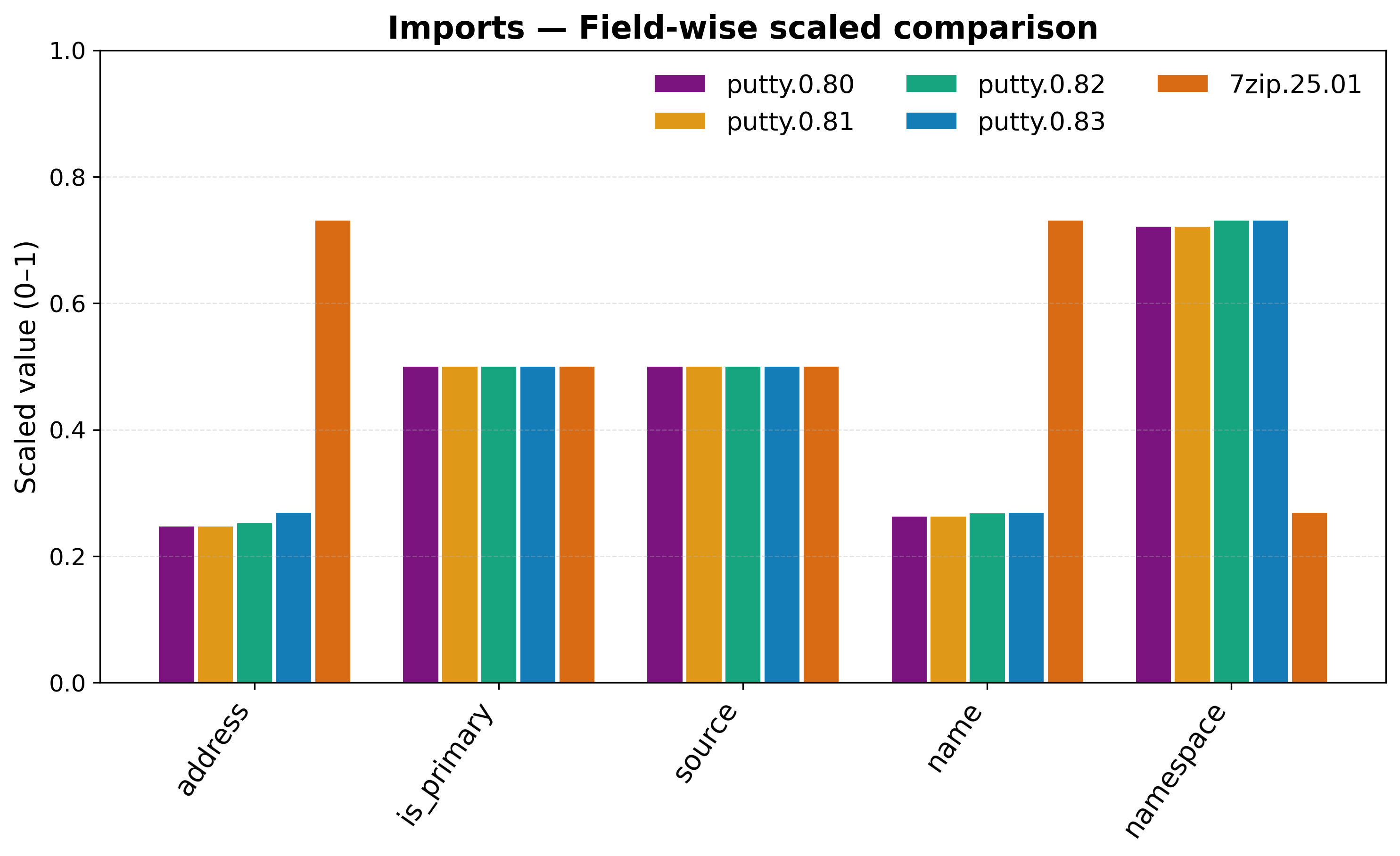}
	\caption{Field-wise import comparison: PuTTY (4 versions) vs. 7-Zip.}
	\label{fig:bars-imports}
\end{figure}

\begin{figure}[!ht]
	\centering
	\includegraphics[width=0.9\linewidth]{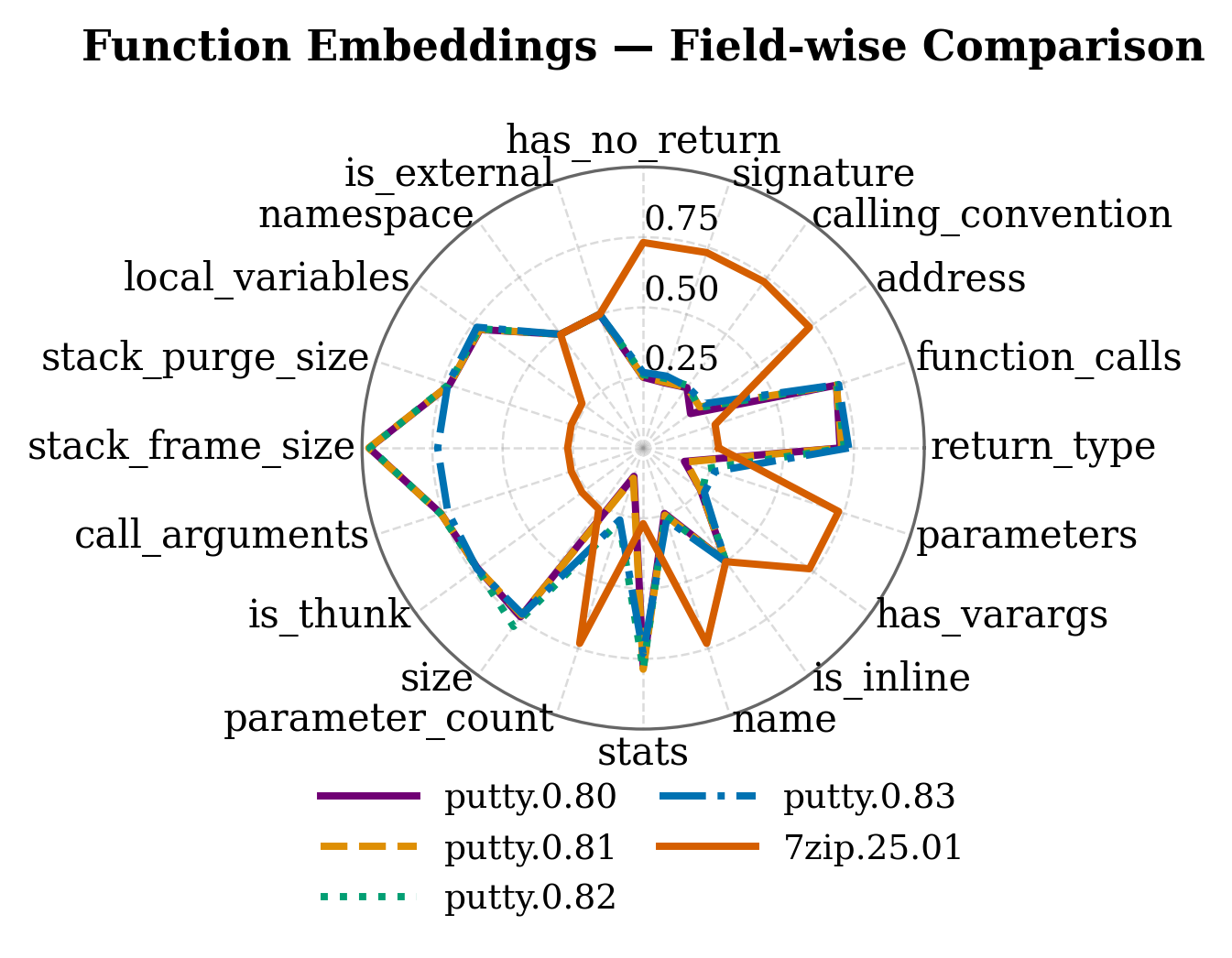}
	\caption{Field-wise function embeddings comparison (cosine similarity): PuTTY (4 versions) vs. 7-Zip.}
	\label{fig:radar-functions}
\end{figure}

\begin{figure*}[!t]
	\centering
	\includegraphics[width=0.9\textwidth]{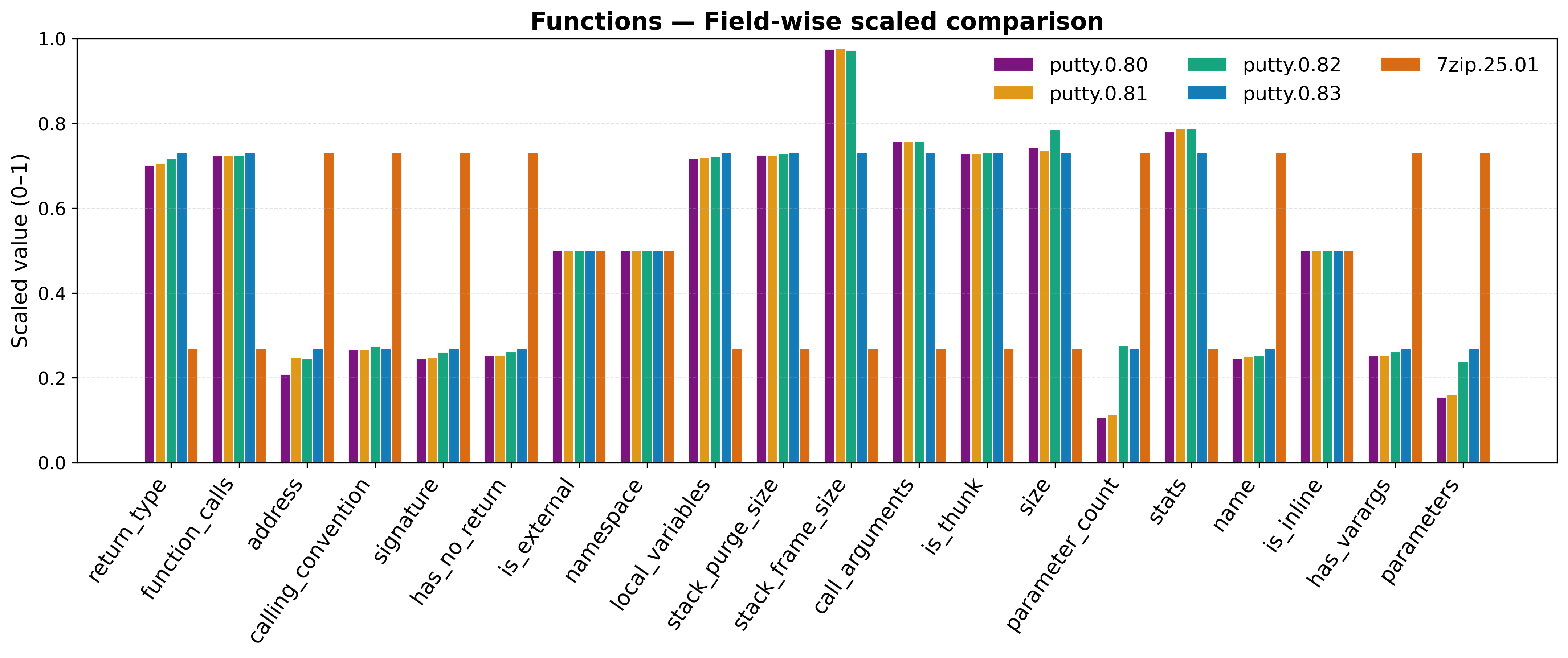}
	\caption{Field-wise function comparison: PuTTY (4 versions) vs. 7-Zip.}
	\label{fig:bars-functions}
\end{figure*}

We then analyze functions, where PuTTY versions expand along structural descriptors such as locals, call connectivity, and size, while 7-Zip broadens on ABI/address-facing cues including parameters, calling conventions, and varargs—patterns consistent with wrapper or thunk layers leading into optimized cores (see Fig.~\ref{fig:radar-functions} and Fig.~\ref{fig:bars-functions}).

\begin{figure}[!ht]
	\centering
	\includegraphics[width=0.9\linewidth]{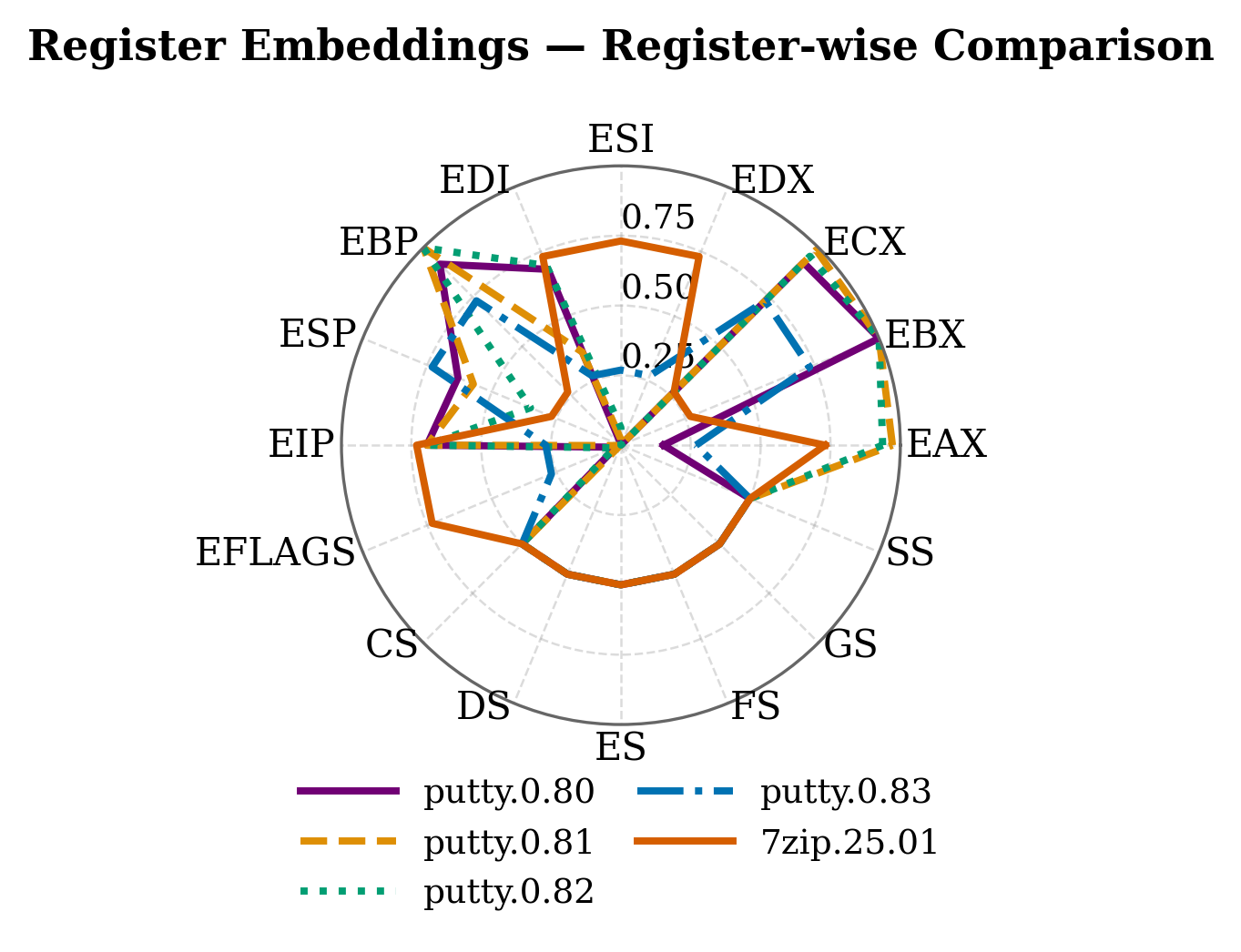}
	\caption{Register-wise embeddings comparison (cosine similarity): PuTTY (4 versions) vs. 7-Zip.}
	\label{fig:radar-registers}
\end{figure}

\begin{figure}[!ht]
	\centering
	\includegraphics[width=0.9\linewidth]{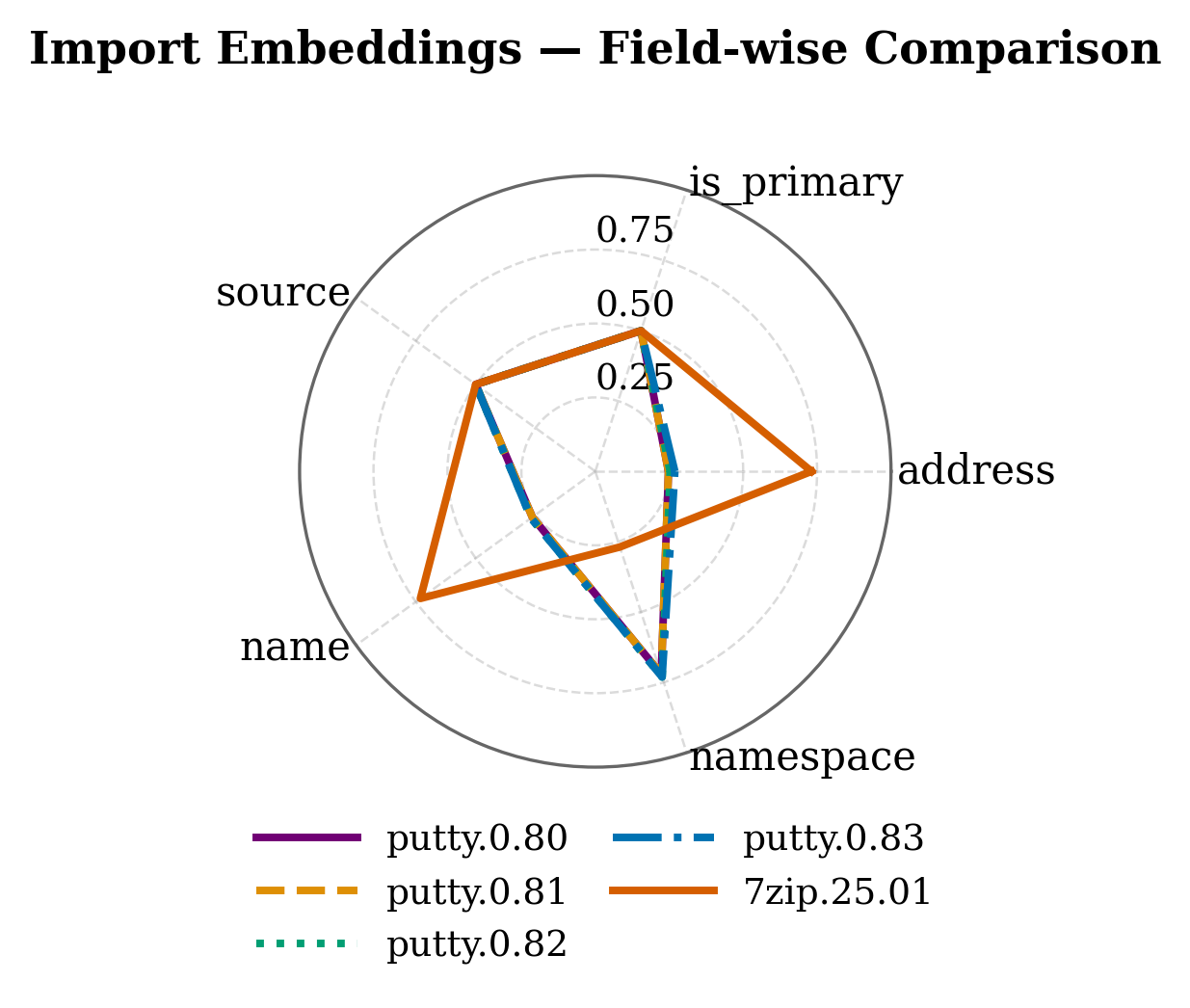}
	\caption{Field-wise import embeddings comparison (cosine similarity): PuTTY (4 versions) vs. 7-Zip.}
	\label{fig:radar-imports}
\end{figure}

\begin{figure}[!ht]
	\centering
	\includegraphics[width=0.9\linewidth]{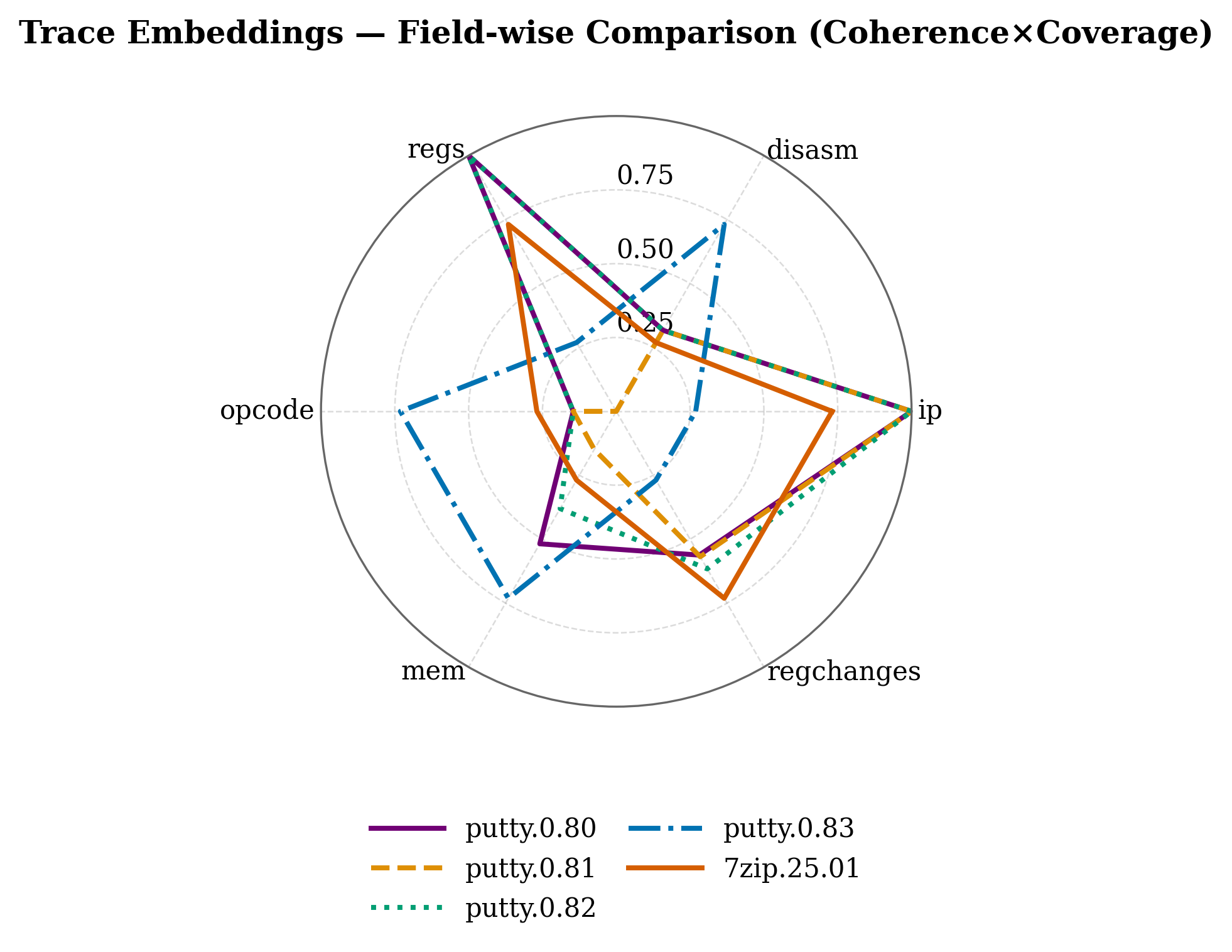}
	\caption{Field-wise trace embeddings comparison (coherence × coverage): PuTTY (4 versions) vs. 7-Zip.}
	\label{fig:radar-traces}
\end{figure}

Imports reinforce this asymmetry; we find that 7-Zip again privileges name/address while PuTTY retains stronger namespace contributions, although is\_primary and source remain near parity  (see Fig.~\ref{fig:bars-imports} and Fig.~\ref{fig:radar-imports}).

\begin{figure}[!t]
	\centering
	\includegraphics[width=0.9\linewidth]{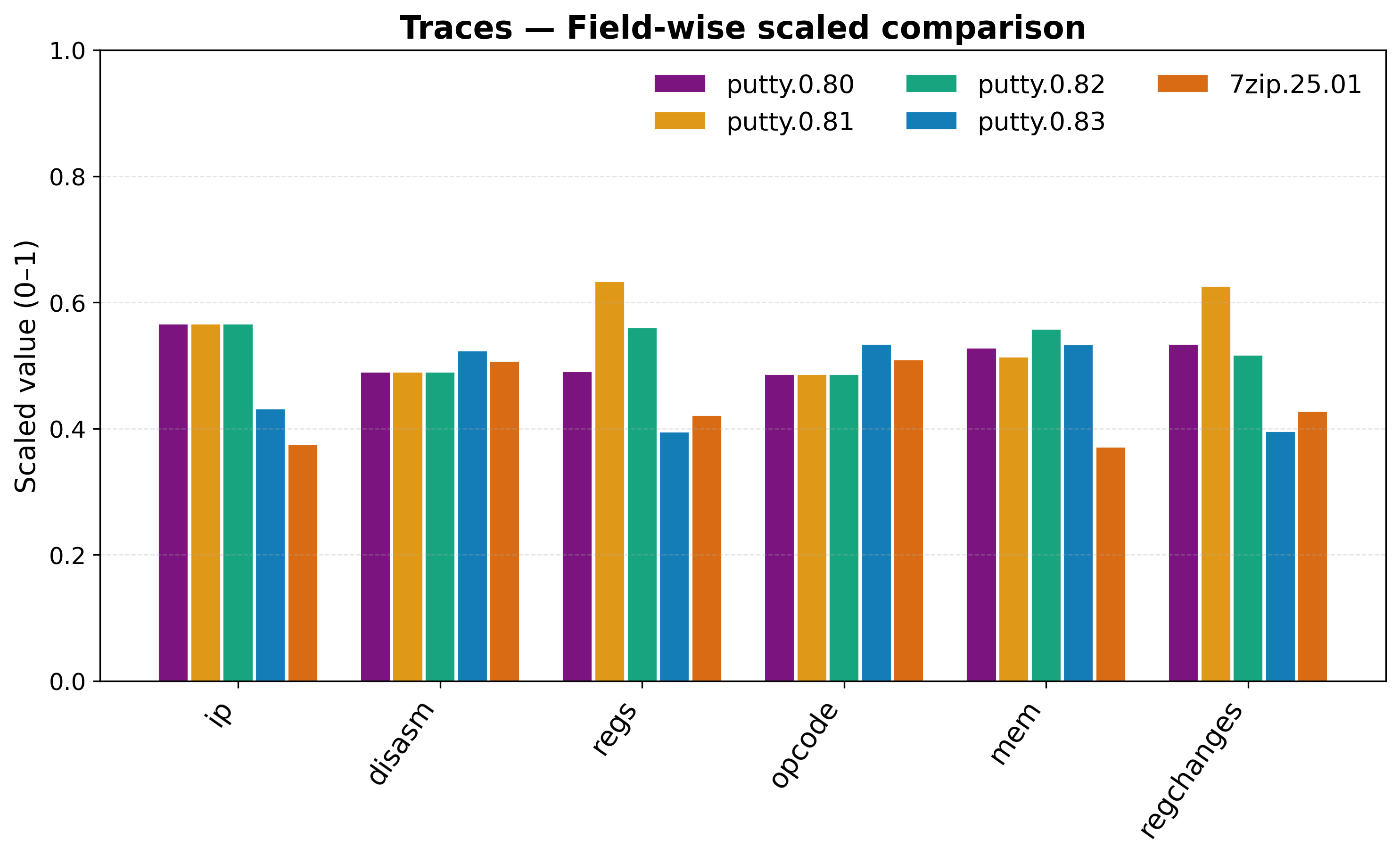}
	\caption{Field-wise dynamic trace comparison: PuTTY (4 versions) vs. 7-Zip.}
	\label{fig:bars-traces}
\end{figure}

\begin{figure*}[!t]
	\centering
	\includegraphics[width=0.9\textwidth]{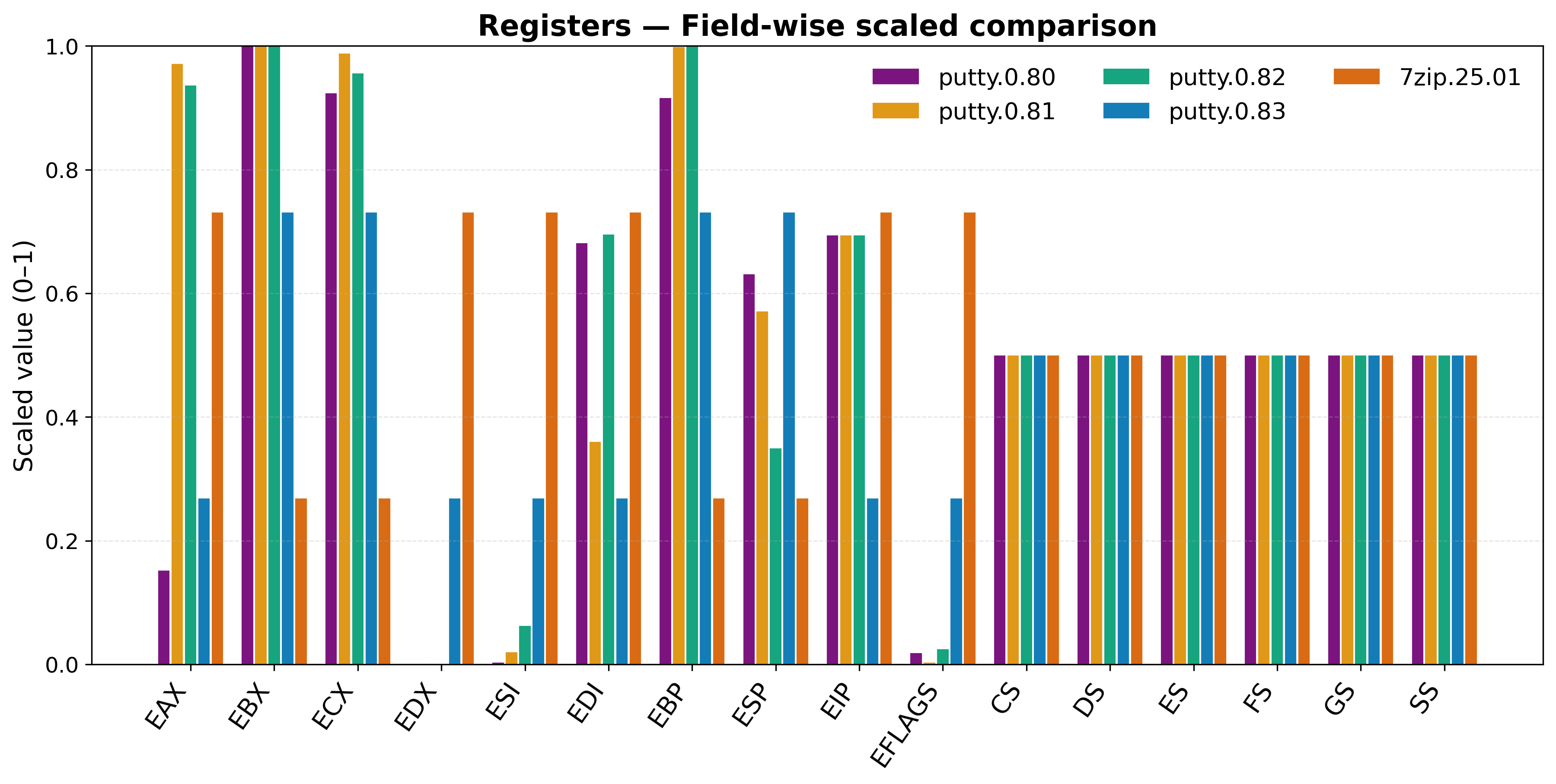}
	\caption{Register-wise comparison: PuTTY (4 versions) vs. 7-Zip.}
	\label{fig:bars-registers}
\end{figure*}

At the register level, we observe PuTTY versions tilting toward frame and control-flow bookkeeping registers (EBP, EBX, EIP, ESP, ECX), while 7-Zip concentrates on accumulator and source-index registers (EAX, ESI), consistent with register-tight inner loops (see Fig.~\ref{fig:radar-registers} and Fig.~\ref{fig:bars-registers}).

Finally, dynamic traces Fig.~\ref{fig:radar-traces} complete the picture; we see PuTTY versions exhibiting broader instruction and memory activity, whereas 7-Zip shows denser register churn and tighter IP locality, reinforcing the narrative derived from functions and imports.

We complement these qualitative comparisons with quantitative similarity scores. We use cosine similarity to measure alignment between embedding vectors: for two nonzero \\ vectors $\mathbf{x},\mathbf{y}\in\mathbb{R}^d$, we compute
\begin{equation}
	\cos(\mathbf{x}, \mathbf{y}) = \frac{\mathbf{x}^\top \mathbf{y}}{\lVert \mathbf{x} \rVert_2 \, \lVert \mathbf{y} \rVert_2}.
	\label{eq:cosine}
\end{equation}
This measure is widely used in information retrieval and representation learning~\cite{b76}. The feature-level cosines skew strongly negative, indicating highly divergent representations across views, consistent with the differences we observe in (see Fig.~\ref{fig:bars-exports} and Fig.~\ref{fig:bars-traces}).

\section{Conclusion}
    This work introduced a novel methodology for detecting and analyzing similarities between program artifacts by integrating both static and dynamic perspectives. Unlike many existing approaches that primarily rely on static analysis, our framework incorporates runtime characteristics such as CPU usage, imports, exports, and register memory allocation. This integration enables a comprehensive multi-view representation of the analyzed artifcats. Furthermore, the methodology enhances interpretability by providing human-understandable insights into the underlying similarity signals. In doing so, it lays a solid foundation for advancing explainable AI (xAI) techniques in binary analysis and opens new directions for more transparent and reliable program comparison in future research.
    
    From a research perspective, future work will focus on developing a comprehensive database encompassing a diverse set of binaries, each enriched with both static and dynamic feature annotations derived from our analysis. This extension aims to facilitate broader and more systematic evaluations. However, key challenges remain, including the dependence on the availability and completeness of dynamic traces, potential environmental biases inherent in their collection, and the increased computational cost associated with aggregating and reconciling multiple execution runs for programs with varied inputs.

% Include all entries from references.bib, even if not cited explicitly

\end{document}